# KH-PINN: Physics-informed neural networks for Kelvin-Helmholtz instability with spatiotemporal and magnitude multiscale

Jiahao Wu, Yuxin Wu[*], Xin Li, Guihua Zhang

*Department of Energy and Power Engineering, Key Laboratory for Thermal Science and Power Engineering of Ministry of Education, Tsinghua University, Beijing 100084, China.*

(* Corresponding author. Tel: 010-62799641. Email: wuyx09@tsinghua.edu.cn)

**Abstract:** Prediction of Kelvin-Helmholtz instability (KHI) is crucial across various fields, requiring extensive high-fidelity data. However, experimental data are often sparse and noisy, while simulated data may lack credibility due to discrepancies with real-world configurations and parameters. This underscores the need for field reconstruction and parameter inference from sparse, noisy data, which constitutes inverse problems. Based on the physics-informed neural networks (PINNs), the KH-PINN framework is established in this work to solve the inverse problems of KHI flows. By incorporating the governing physical equations, KH-PINN reconstructs continuous flow fields and infer unknown transport parameters from sparse, noisy observed data. The 2D unsteady incompressible flows with both constant and variable densities are studied. To our knowledge, this is the first application of PINNs to unsteady incompressible flows with variable densities. To address the spatiotemporal multiscale issue and enhance the reconstruction accuracy of small-scale structures, the multiscale embedding (ME) strategy is adopted. To address the magnitude multiscale issue and enhance the reconstruction accuracy of small-magnitude velocities, which are critical for KHI problems, the small-velocity amplification (SVA) strategy is proposed. The results demonstrate that KH-PINN can accurately reconstruct the fields with complex, evolving vortices and infer unknown parameters across a broad range of Reynolds numbers. Additionally, the energy-decaying and entropy-increasing curves are accurately obtained. The effectiveness of ME and SVA is validated through comparative studies, and the anti-noise and few-shot learning capabilities of KH-PINN are also validated. The code for this work is available at https://github.com/CAME-THU/KH-PINN.

## 1. Introduction

Kelvin-Helmholtz instability (KHI) is a fluid instability that arises from velocity shear within a single continuous fluid or from a velocity difference at the interface between two fluids. KHI is ubiquitous in nature, manifesting in phenomena such as cloud formations, surface ripples on lakes, and exhaled breath, and it frequently occurs in various industrial facilities, particularly in the jet flows used in internal combustion engines, aero engines, gas turbines, and jet cooling systems [1]. Thus, KHI is a critical issue across various fields, including astronomy [2, 3], meteorology [4, 5], oceanography [6, 7], life science [8], environmental science, and engineering [9]. From a fluid physics perspective, KHI typically describes the mixing processes in stratified flows and is a key mechanism for laminar-turbulent transition in free shear flows [10-12]. In engineering contexts, the necessity of predicting KHI is highlighted in many scenarios , such as unstable jet flames caused by KHI, which can jeopardize the safe operation of industrial facilities [13-15]. However, developing KHI prediction models requires substantial of high-fidelity data, which is sparse and noisy in experimental settings. Although numerical simulation can mitigate this limitation, they frequently lack credibility due to discrepancies between simulated and real-world configurations and parameters. Therefore, it is necessary to develop methods for high-resolution reconstruction and parameter inference of KHI based on sparse data, forming the foundation for constructing reliable KHI prediction models.

Data assimilation effectively addresses the sparse-data issue by integrating limited experimental data into numerical models, yielding more detailed flow information, including flow fields and unknown parameters [16]. This approach enables mutual enhancement between experiments and simulations: numerical simulations refine experimental data resolution, while experimental data improve simulation credibility. In recent years, physics-informed neural networks (PINNs) have emerged as a key method for seamlessly integrating neural network (NN) learning with physical laws represented as partial differential equations (PDEs) [17-19]. PINNs can solve both forward PDE problems, where the PDE systems are well-posed with known parameters, boundary conditions (BCs), and initial conditions (ICs), and inverse PDE problems, where parameters, BCs, and ICs may be incomplete but some solution data are available, termed observation conditions (OCs) in this paper. The observed data can be sparse and noisy, simulating practical experimental scenarios, thus aligning inverse PDE problems with data assimilation. In PINNs, an NN represents the PDE solution, and auto-differentiation (AD) computes the PDE residuals that need to be minimized. The meshless nature, data integration capability, and zero discretization error of AD are key advantages of PINNs over traditional numerical methods for solving PDE problems.

Numerous studies have applied PINNs to inverse problems of fluid flows [20-36]. In an early work, Rassi et al. [33] used PINNs to reconstruct velocity and pressure fields from observed passive scalar data. The algorithm is highly flexible, capable of being implemented on arbitrary domains and geometries. Wu et al. [27] proposed the variable linear transformation improved PINN (VLT-PINN) to solve both forward and inverse problems of thin-layer flows, including the jets, wakes, mixing layers, and boundary layers. All of these are incompressible flows, and studies on compressible flows have also been conducted [32, 34, 35, 37]. For example, Jagtap et al. [35] applied the extended PINN (XPINN) to reconstruct supersonic flows from the density gradient data, subject to the constraints of Euler equations and entropy conditions. In addition to the standard Navier-Stokes (NS) equations, averaged NS equations have been employed for inverse problems as well [28-30]. Von Saldern et al. [30] applied PINNs to turbulent mean flow assimilation of jet flows based on reduced equations and simplified models, while Luo et al. [29] leveraged PINNs to infer and optimize five parameters in the $k$-$\varepsilon$ model using direct numerical simulation (DNS) data. The application of PINNs to reacting flows, though a highly active research area, remains challenging due to the inherent nonlinearity. Sitte et al. [25], Wang et al. [24], Liu et al. [26], and Wu et al. [36] used PINNs to reconstruct flow fields of puffing fires, rotating detonation combustor, 3D turbulent flames, and 1D counterflow flames, respectively. These studies predominantly utilized high-fidelity simulated data, which serves as a foundation for practical applications of PINNs but still diverges significantly from real-world conditions. Consequently, some studies have explored the feasibility of employing PINNs for inverse problems using experimental data [20, 22, 23, 30]. Cai et al. [23] used PINNs to infer 3D velocity and pressure fields of the flow over an espresso cup using only the temperature data from tomographic background oriented Schlieren (Tomo-BOS), and Wang et al. [22] applied PINNs to reconstruct dense velocity fields of the wake flow behind a hemisphere using data from tomographic particle image velocimetry (Tomo-PIV).

In KHI flows, vortices that formed at the initial shear layers continuously wrap the fluid into increasingly thinner filaments, resulting in a distinctly multiscale flow. Due to the spectral bias or frequency principle of NNs [38-41], NN-based models often struggle to capture high-frequency/small-scale features, making multiscale problems particularly challenging for PINNs. Nevertheless, numerous studies have sought to mitigate this issue [42-54]. Early research identified a simple but effective solution: multiscale embedding (ME), which can be implemented both linearly and nonlinearly. For linear ME, Liu et al. [52] first proposed the MscaleDNN to solve the Poisson-Boltzmann equation, and this approach has since been applied to various other problems [45, 51]. For nonlinear ME, Fourier ME has proven effective [42], which is now the most adopted ME for PINNs solving multiscale problems. Wang et al. [54] analyzed the spectral bias of NNs through the lens of neural tangent kernel (NTK) theory, demonstrating that PINNs tend to learn solutions and partial differential equation (PDE) residuals at the lowest frequencies first. Building on this, they developed a multiscale PINN that embeds spatiotemporal multiscale random Fourier features, yielding robust and accurate results for a range of multiscale problems. This random Fourier ME has since been utilized in numerous studies involving PINNs for multiscale issues [43, 45, 55]. Additionally, some other methods have been developed, such as utilizing homogenization theory [49], modifying the loss function [43], and applying transfer learning [44].

Despite the advancements in the aforementioned studies, several gaps remain concerning our problem. (1) There is a notable scarcity of studies employing PINNs to solve KHI and transitional flow problems, which exhibit both laminar and turbulent characteristics. Most studies focus exclusively on either pure laminar or pure turbulent flows. While some studies applied PINNs to Rayleigh-Taylor instabilities (RTI) [56-58], this instability differs fundamentally from KHI. Although Hanrahan et al. [59] used PINNs to predict transitional flow around a turbine blade, the turbulent transition in wall-bounded shear flows are distinct from those in free shear flows. While many studies used NNs for surrogate and transition modeling of KHI [60-65], these approaches are purely data-driven, lack physical constraints, and are not designed for inverse problems. Mishra et al. [66] applied PINNs to solve the forward problem of KHI but used the inviscid Euler equations rather than the full NS equations. (2) There is a lack of studies investigating PINNs for solving incompressible flows with variable densities. Most research has concentrated on incompressible flows with constant density or on compressible flows. In fact, even traditional numerical methods have rarely been applied to simulate such flows [67-72]. Incompressible flows with variable densities occur when multiple incompressible, fully miscible fluids with different densities are mixed, representing a type of multi-species flow with broad applications and necessitating further exploration. (3) While PINNs have been extensively explored for spatiotemporal multiscale problems, magnitude multiscale problems have received limited attention. Spatiotemporal multiscale pertains to the distribution of variables over time and space, while magnitude multiscale relates to the order of magnitude of these variables. They can be simplified as multi-frequency and multi-magnitude, respectively. KHI is typically triggered by minor-magnitude perturbations, which are crucial for field reconstruction but challenging for conventional PINNs to capture, as minor-magnitude contributions often receive insufficient focus from the network. Due to the magnitude multiscale, common normalization of minor-magnitude variables can disrupt the overall normalization of variables, potentially leading to gradient explosion and divergence during training.

In this work, to investigate the applicability of PINNs for field reconstruction and parameter inference of KHI, we develop the KH-PINN model for solving inverse problems of KHI based on PINNs. To broaden the scope of applications, both 2D incompressible flows with constant and variable densities are examined. To mitigate the difficulties posed by spatiotemporal multiscale and enhance the reconstruction capabilities of PINNs for small-scale structures, the nonlinear ME is adopted, alleviating the spectral bias inherent in NNs. Furthermore, to mitigate the difficulties posed by magnitude multiscale and enhance the reconstruction capabilities of PINNs for small-magnitude velocities, we propose the small-velocity amplification (SVA) strategy. This strategy continuously and nonlinearly transforms the velocities, preserving the normal-magnitude components while amplifying the small-magnitude ones. The results show that KH-PINN can accurately reconstruct flow fields and infer unknown transport parameters across a wide range of Reynolds numbers, using sparse data obtained from high-fidelity direct numerical simulation (DNS). The ME and SVA strategies are shown to effectively improve the reconstruction accuracy for small-scale structures and small-magnitude velocities, respectively. Moreover, the anti-noise and few-shot learning capabilities of KH-PINN are validated. We also highlight the advantages of KH-PINN over traditional data fitting methods using NNs, as well as the benefits of the employed NN architecture compared to conventional ones, which will be discussed in detail in the subsequent sections.

In Section 2, the methodology is first introduced, including the problem definition, reference solution, and the KH-PINN framework. Section 3 presents and discusses the results, beginning with an overview of the overall outcomes, followed by the validation of the proposed strategies and configurations. Finally, Section 4 provides the conclusions of the paper.

## 2. Methodology

This section begins by describing how the reference solutions are obtained, including the governing equations, problem configurations, and key features of the solution. Then, the details of KH-PINN for solving the inverse problems of KHI are presented, outlining its fundamental principles and adopted strategies. Finally, the PINN configurations and implementation details are provided.

### 2.1 Problem description and reference solution

Both the 2D incompressible KHI problems with variable and constant densities are governed by the incompressible Navier-Stokes (NS) equations:

$$\begin{cases} \nabla \cdot \mathbf{U} = 0 \\ \dfrac{\partial \rho}{\partial t} + \mathbf{U} \cdot \nabla \rho = D \Delta \rho \\ \dfrac{\partial \mathbf{U}}{\partial t} + \mathbf{U} \cdot \nabla \mathbf{U} = -\dfrac{1}{\rho} \nabla p + \dfrac{1}{\rho} \nabla \cdot \left[ \rho \nu \left( \nabla \mathbf{U} + \nabla \mathbf{U}^T \right) \right] \\ \dfrac{\partial c}{\partial t} + \mathbf{U} \cdot \nabla c = \dfrac{1}{\rho} \nabla \cdot \left( \rho D \nabla c \right) \end{cases} \quad (1)$$

where $\mathbf{U} = [u, v]$ is the velocity vector, $\rho$ is the density, $p$ is the pressure, and $c$ is the passive scalar, which can represent the mass fraction of the dye or other species. The $\nu$ is the viscosity and $D$ is the diffusion coefficient of the passive scalar, which are both considered as constants. The first equation represents the condition of incompressibility and the last three equations describe the transport equations of mass, momentum, and passive scalar, respectively. In miscible multi-species flows, not only the species but also the total mass exhibit diffusive motions, so a diffusive term is added to the original single-species mass conservation equation. For simplicity, the diffusion coefficient is set equal to that of $c$. In homogeneous flows, the density is spatially constant, allowing the governing equations to be simplified to:

$$\begin{cases} \nabla \cdot \mathbf{U} = 0 \\ \dfrac{\partial \mathbf{U}}{\partial t} + \mathbf{U} \cdot \nabla \mathbf{U} = -\dfrac{1}{\rho} \nabla p + \nu \Delta \mathbf{U} \\ \dfrac{\partial c}{\partial t} + \mathbf{U} \cdot \nabla c = D \Delta c \end{cases} \quad (2)$$

which is a more common form of the incompressible NS equations.

DNS was implemented for the KHI problem to get its reference solutions. The calculation domain is: $x \in [0, 0.5]$ m, $y \in [0, 1]$ m, $t \in [0, 5]$ s. Periodic BCs are assumed in both directions. The ICs are:

$$\begin{cases} \rho_{t=0} = 1 + \dfrac{1}{2}\dfrac{\delta\rho}{\rho_0}\left[\tanh\left(\dfrac{y-y_1}{a}\right) - \tanh\left(\dfrac{y-y_2}{a}\right)\right] \\ u_{t=0} = u_{\text{flow}}\left[\tanh\left(\dfrac{y-y_1}{a}\right) - \tanh\left(\dfrac{y-y_2}{a}\right) - 1\right] \\ v_{t=0} = A\sin(4\pi x) \\ p_{t=0} = p_0 \\ c_{t=0} = \dfrac{1}{2}\left[\tanh\left(\dfrac{y-y_1}{a}\right) - \tanh\left(\dfrac{y-y_2}{a}\right)\right] \end{cases} \quad (3)$$

where the IC of density is only valid for variable density cases. For constant density cases, $\rho = 1$ kg/m$^3$. The double tangent functions make the corresponding fields platform functions, with the shear layers forming at the platform edges. The positions and smoothness of the two shear layers are controlled by $a$ and $y_i$, respectively. If $v_{t=0}$ is zero, the flow becomes a diffusive stratified flow characterized by progressively wider mixing layers, where vortices do not occur. To generate the KHI, a small initial perturbation produced by the velocity $v$ is required, which is modeled as a sinusoidal function in this work. The meanings and values of the parameters are summarized in Table 1. The Reynolds numbers we tested are $Re = [1000, 2000, \ldots, 10000]$. The viscosity $\nu = 1/Re$ and we set $D = \nu$ for simplicity.

**Table 1**

Parameters of the initial conditions, using SI units.

| Parameter | Symbol | Value |
|---|---|---|
| Base density | $\rho_0$ | 1 |
| Density jump | $\delta\rho_0$ | 1 |
| Shear layer width | $a$ | 0.025 |
| Shear layer positions | $y_1, y_2$ | 0.25, 0.75 |
| Main flow velocity | $u_{\text{flow}}$ | 1 |
| Perturbation amplitude | $A$ | 0.01 |
| Initial pressure | $p_0$ | 1 |

The DNS was implemented with Dedalus [73], an open-source Python library designed for solving PDEs using spectral methods. The mesh sizes ($N_x \times N_y$) are 512×1024 and 1024×2048 for constant and variable density cases, respectively. For time-stepping, we used a third-order, four-stage Runge-Kutta method (RK443 in Dedalus) with a CFL safety factor of 0.2. Fig. 1 illustrates the turbulent kinetic energy (TKE) spectra at various moments, illustrating the turbulent transition process: Small-scale structures (large $k$) gradually emerge in the early stages, and then the slope of the curve approaches −5/3. The subsequent decrease in the curve indicates the energy-decaying stage. Fig. 2 shows the magnitude of $v$, where two types of small-magnitude velocities are observed: The minor velocity perturbations that initiate the KHI at the beginning and the minor velocities that develop during the late dissipation stage. Fig. 1 and Fig. 2 clearly show the aforementioned spatiotemporal and magnitude multiscale characteristics of KHI, respectively, which pose challenges for PINNs to reconstruct the small-scale structures and small-magnitude velocities, and are thus the focus of our investigation.

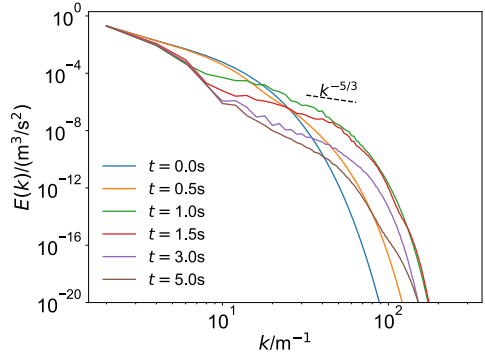

**Fig. 1.** TKE spectra at different moments, where *k* is the wave number. Case: Constant density, *Re* = 10000.

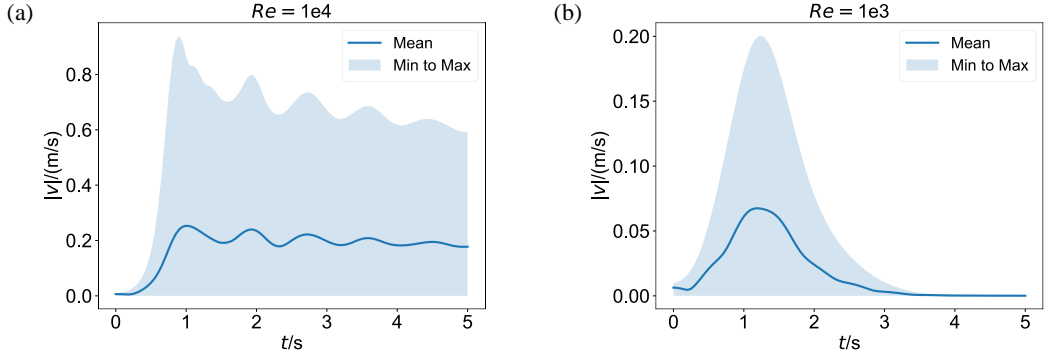

**Fig. 2.** The magnitude of *v* over time. The solid lines represent the mean value and the shaded parts represent the range of *v*. (a) *Re* = 10000. (b) *Re* = 1000. Both are constant density cases.

## 2.2 KH-PINN framework

Fig. 3 presents a schematic of KH-PINN framework for field reconstruction and parameter inference of KHI by solving the inverse PDE problems (the yellow part), which consist of PDEs and OCs, given by

$$\mathcal{F}_i\left[\mathbf{u}(\mathbf{x});\boldsymbol{\kappa}\right]=0, \ \mathbf{x}\in\Omega, \ i=1,2,...,N_{\text{PDE}} \quad (4)$$

$$\mathcal{O}_i\left[\mathbf{u}(\mathbf{x});\boldsymbol{\kappa}\right]=0, \ \mathbf{x}\in\Omega, \ i=1,2,...,N_{\text{OC}} \quad (5)$$

where $\mathcal{F}$ and $\mathcal{O}$ are the corresponding operators, $\Omega$ is the calculation domain, $\mathbf{x} = [x, y, t]$ is the vector of independent variables, and $\mathbf{u}$ is the vector of dependent variables, which is $[\rho, u, v, p, c]$ and $[u, v, p, c]$ for variable and constant density cases, respectively. The vector $\boldsymbol{\kappa}$ represents system parameters, such as the material properties and transport coefficients. Note that the ICs and BCs are absent and $\boldsymbol{\kappa}$ may be unknown, which renders the system ill-posed for forward problems but can be solved for inverse problems since the introduction of OCs, where the observed data may be sparse and partial (that is, not all components of $\mathbf{u}$ are observed). Based on the observations, field reconstruction is to solve the continuous field, while parameter inference is to infer the unknown parameters ($\boldsymbol{\kappa}$).

To solve the PDE problem, KH-PINN learns the solution function $\mathbf{u}(\mathbf{x})$ based on an NN:

$$\begin{aligned} \mathbf{x}^{\#} &= f_{\text{input}}(\mathbf{x}) \\ \mathbf{u}_{\boldsymbol{\theta}}^{\#} &= \text{NN}(\mathbf{x}^{\#};\boldsymbol{\theta}) \\ \mathbf{u}_{\boldsymbol{\theta}} &= f_{\text{output}}(\mathbf{u}_{\boldsymbol{\theta}}^{\#}) \end{aligned} \quad (6)$$

where $\mathbf{x}^{\#}$ and $\mathbf{u}_{\boldsymbol{\theta}}^{\#}$ are the direct input and output of the NN, respectively, and $\boldsymbol{\theta}$ is the trainable parameters of the NN. When the transformation functions (*f*) are normalization or denormalization, the corresponding hash superscripts (#) are replaced with star ones (*). The derivatives of $\mathbf{u}_{\boldsymbol{\theta}}$ with respect to $\mathbf{x}$ can be computed by AD and substituted into $\mathcal{F}$ and $\mathcal{O}$ to get the residuals of the PDEs and OCs. The mean square errors (MSEs) of each residual are taken as the corresponding loss terms:

$$\mathcal{L}_{\text{PDE}i}(\boldsymbol{\theta},\boldsymbol{\kappa};\mathcal{T}_{\text{PDE}i}) = \frac{1}{|\mathcal{T}_{\text{PDE}i}|} \sum_{\mathbf{x}_j \in \mathcal{T}_{\text{PDE}i}} \left| \mathcal{F}_i \left[ \mathbf{u}_{\boldsymbol{\theta}}(\mathbf{x}_j);\boldsymbol{\kappa} \right] \right|^2, \quad \mathcal{T}_{\text{PDE}i} := \{ \mathbf{x}_j \mid \mathbf{x}_j \in \Omega \}_{j=1}^{|\mathcal{T}_{\text{PDE}i}|} \tag{7}$$

$$\mathcal{L}_{\text{OC}i}(\boldsymbol{\theta},\boldsymbol{\kappa};\mathcal{T}_{\text{OC}i}) = \frac{1}{|\mathcal{T}_{\text{OC}i}|} \sum_{\mathbf{x}_j \in \mathcal{T}_{\text{OC}i}} \left| \mathcal{O}_i \left[ \mathbf{u}_{\boldsymbol{\theta}}(\mathbf{x}_j);\boldsymbol{\kappa} \right] \right|^2, \quad \mathcal{T}_{\text{OC}i} := \{ (\mathbf{x}_j, u_{i,j}) \mid \mathbf{x}_j \in \Omega \}_{j=1}^{|\mathcal{T}_{\text{OC}i}|} \tag{8}$$

where $\mathcal{T}_{\text{PDE}i}$ is the coordinate point set sampled on the domain and $\mathcal{T}_{\text{OC}i}$ is the observed point set for component $u_i$. The total loss function is the weighted sum of all loss terms:

$$\mathcal{L}(\boldsymbol{\theta},\boldsymbol{\kappa}) = \sum_{i=1}^{N_{\text{PDE}}+N_{\text{OC}}} w_i \mathcal{L}_i(\boldsymbol{\theta},\boldsymbol{\kappa}) \tag{9}$$

The trainable parameters, including NN parameters and unknown system parameters, are updated via gradient-descent algorithms:

$$\boldsymbol{\theta}_{k+1} = \boldsymbol{\theta}_k - \eta_k f_{\text{opt}}(\nabla_{\boldsymbol{\theta}} \mathcal{L}_k) \tag{10}$$

$$\boldsymbol{\kappa}_{k+1} = \boldsymbol{\kappa}_k - \eta_k f_{\text{opt}}(\nabla_{\boldsymbol{\kappa}} \mathcal{L}_k) \tag{11}$$

where $k$ is the index of the current iteration, $\eta_k$ is the learning rate, and $f_{\text{opt}}$ is a function specified by the optimizer. The learning process will stop after predefined iterations/epochs or the convergence conditions are satisfied.

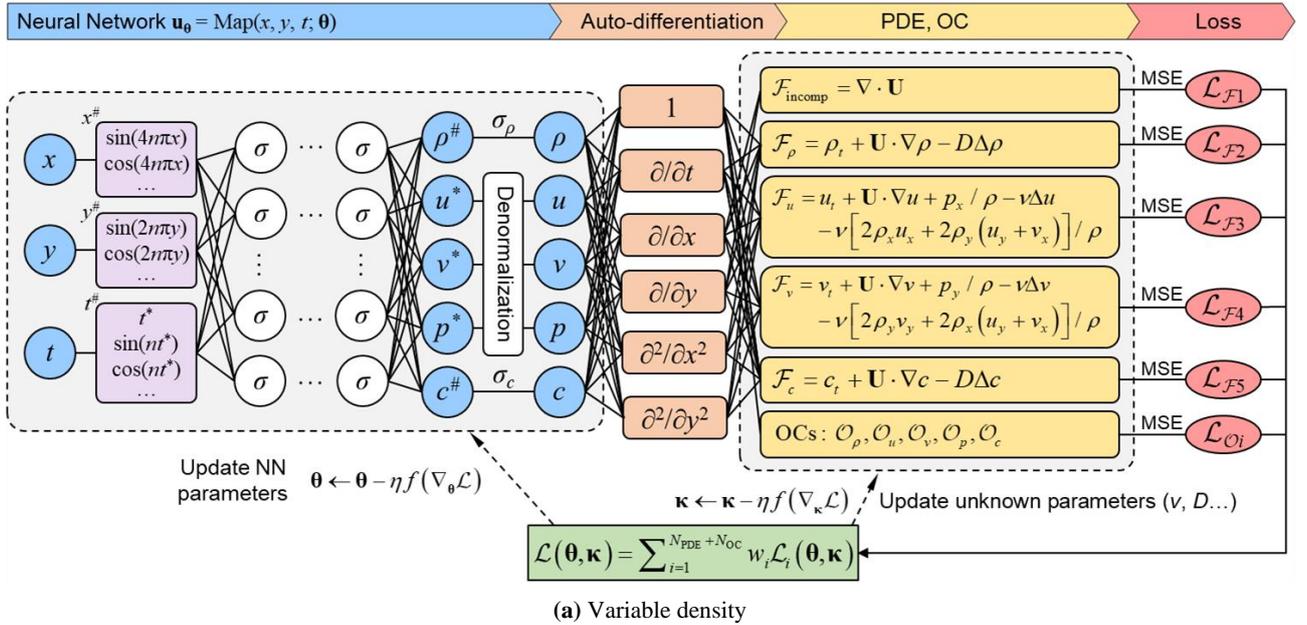

(a) Variable density

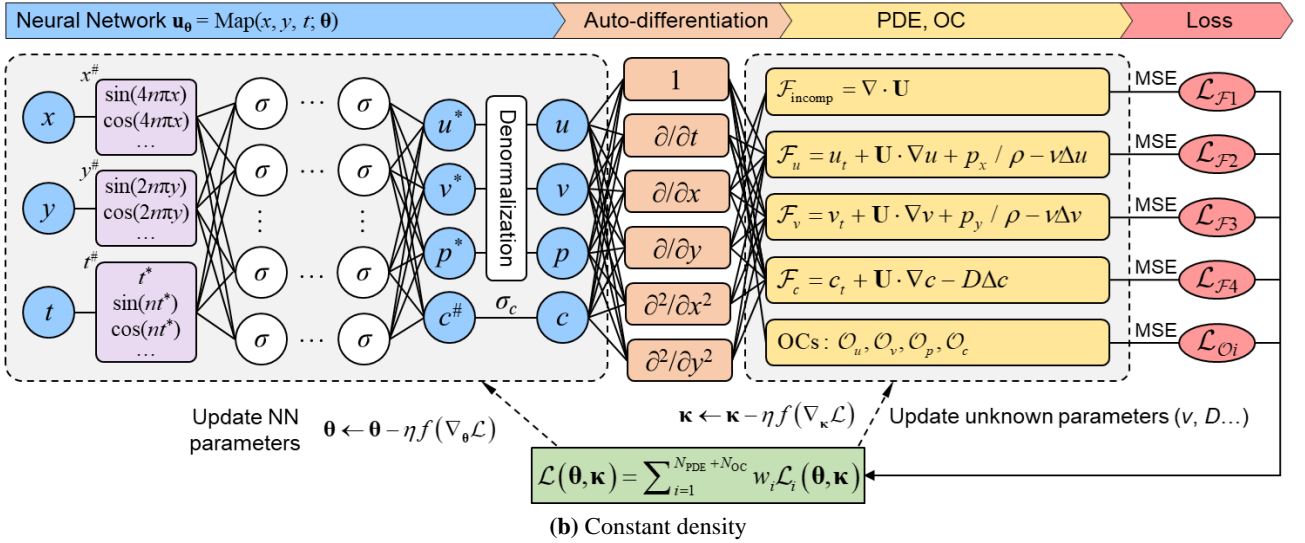

(b) Constant density

**Fig. 3.** Schematic of KH-PINN framework for solving inverse problems of KHI. OC: observation condition. The star superscript represents normalized variables. In practice, the PDEs and OCs of normalized variables instead of the original ones are adopted.

## 2.3 Transformation of variables

For robust training, it is essential to apply appropriate transformations to the variables, including the input and output variables, as well as the unknown parameters to be inferred. To address the challenges in training KH-PINN arising from the multiscale nature of KHI, we have designed both linear and nonlinear transformations. To avoid gradient vanishing and explosion in training NNs, the linear transformations are used for normalization:

$$\varphi^* = \alpha_\varphi (\varphi + \delta_\varphi), \quad \varphi = x, y, t, \rho, u, v, p, c, \nu, D \tag{12}$$

which is the same as in our previous work [27], and we set $\delta_\varphi = 0$ for $\rho$, $u$, $v$, $p$, $c$, $\nu$, and $D$. Note that not all normalized variables are used in KH-PINN, as Fig. 3 shows, but the scaling factors ($\alpha$) are used for the normalization of PDEs and OCs in practical training (not shown in Fig. 3). Specifically:

$$\mathcal{F}^*_{\text{incomp}} = \mathcal{F}_{\text{incomp}}, \quad \mathcal{F}^*_\varphi = \frac{\alpha_\varphi}{\alpha_t} \mathcal{F}_\varphi, \quad \varphi = \rho, u, v, c \tag{13}$$

$$\mathcal{O}^*_\varphi = \alpha_\varphi \mathcal{O}_\varphi, \quad \varphi = \rho, u, v, p, c \tag{14}$$

where the PDEs are normalized based on the main terms (transient terms), according to our previous work [27]. It is easy to derive that solving $\mathcal{F}^*$ and $\mathcal{O}^*$ is equivalent to solving the PDEs and OCs of normalized variables (or nondimensional variables in more studies), but the tedious derivations of the PDEs of the normalized variables can be circumvented by computing the residuals of the original PDEs using AD and multiplying them by the corresponding scaling factors. We set $\alpha_t = 1$ and $\delta_t = -2.5$, ensuring that the normalized $t^*$ is in $[-2.5, 2.5]$. The scaling factors for the dependent variables are defined as the reciprocals of the mean absolute value of the observed data:

$$\alpha_\varphi = \frac{N_{\text{ob}}}{\sum_{i=1}^{N_{\text{ob}}} |\varphi_i|}, \quad \varphi = \rho, u, v, p, c \tag{15}$$

Note that scaling the inferred parameters, which is equivalent to adjusting their learning rates [27], is also necessary to address the difficulty posed by parameters with excessively large or small magnitudes. The scaling factors of $\nu$ and $D$ are both set to 1000, allowing them to be scaled to approximately $\mathcal{O}(0.1)$ [27].

Nonlinear transformations are also applied to both input and output variables, as described below.

### 2.3.1 Input transformation: Multiscale embedding

As previously stated, given the frequency principle (or spectral bias) of NNs, the spatiotemporal multiscale behavior of various physical problems is a major challenge for PINNs to address, which is also the case of KHI flows. To tackle this issue, multiscale embedding is a common strategy (as mentioned in the introduction) and is also adopted in our work. Specifically, the spatiotemporal variables are transformed into multiscale features (as Fig. 3 shows):

$$x^{\#} = \left[\sin(4\pi x), \cos(4\pi x), \sin(8\pi x), \cos(8\pi x), ..., \sin(4n\pi x), \cos(4n\pi x)\right]$$
$$y^{\#} = \left[\sin(2\pi y), \cos(2\pi y), \sin(4\pi y), \cos(4\pi y), ..., \sin(2n\pi y), \cos(2n\pi y)\right] \quad (16)$$
$$t^{\#} = \left[t^*, \sin(t^*), \cos(t^*), \sin(2t^*), \cos(2t^*), ..., \sin(nt^*), \cos(nt^*)\right]$$

Then $\mathbf{x}^{\#} = [x^{\#}, y^{\#}, t^{\#}]$ is input to the NN. The default value of $n$ is 4 in our work.

Some notes are as follows. (1) Our embedding is both a multiscale embedding and a spatially periodic embedding, ensuring that the periodic BCs are precisely satisfied. Specifically, since the domain lengths in the $x$ and $y$-dimensions are 0.5 and 1, respectively, the minimal positive periods of $x^{\#}$ and $y^{\#}$ are also 0.5 and 1, respectively. (2) Unlike the linear multiscale embedding in the original MscaleDNN [52], we adopted the Fourier embedding because it is better suited to multi-frequency problems and can be designed to satisfy the periodic BCs. (3) For the same reason, we did not use the random Fourier embedding as described in Ref. [54], as this would violate the periodic BCs. (4) Retaining the linear term ($t^*$) in $t^{\#}$ is necessary because there are no periodic conditions in the time dimension.

*2.3.2 Output transformation: Small-velocity amplification*

Given the known ranges of $\rho$ and $c$, we applied nonlinear transformations to the NN outputs to ensure that the range constraints were automatically satisfied, thereby preventing the occurrence of unphysical values. Specifically,

$$\rho = \text{Sigmoid}(\rho^{\#}) \cdot \delta\rho / \rho_0 + 1$$
$$c = \text{Sigmoid}(c^{\#}) \quad (17)$$

where $\text{Sigmoid}(x) = (1+e^{-x})^{-1}$.

As previously noted, KHI is typically induced by minor perturbations, which are important for field reconstruction but challenging for conventional PINNs to address. To tackle this issue, we propose a nonlinear transformation strategy to enhance the effectiveness of OCs by enlarging the small magnitude components of the variables while keeping the other components nearly unchanged. Since the KHI is induced by velocity perturbations in our work, the transformations are only applied to $u$ and $v$. Mathematically, the operators of the OCs are defined as:

$$\mathcal{O}_{\varphi} = \varphi_0 - \varphi, \quad \varphi = \rho, p, c \quad (18)$$
$$\mathcal{O}_{\varphi} = f_{\mathbf{U}}(\varphi_0) - f_{\mathbf{U}}(\varphi), \quad \varphi = u, v \quad (19)$$

where $f_{\mathbf{U}}$ is the transformation function. To accomplish the aforementioned purpose, we designed 4 types of functions:

$$f_{\mathbf{U},1}(x) = x + \frac{\arctan((a_1+9)s_1 x) - \arctan(a_1 s_1 x)}{s_1}$$
$$f_{\mathbf{U},2}(x) = x + \frac{9x}{1+s_2 x^2}$$
$$f_{\mathbf{U},3}(x) = x + \frac{\tanh((a_3+9)s_3 x) - \tanh(a_3 s_3 x)}{s_3} \quad (20)$$
$$f_{\mathbf{U},4}(x) = \frac{x}{0.45\tanh(10(|x|-0.25)) + 0.55}$$

where $a$ and $s$ are adjustable parameters and their default values are: $a_1 = 6$, $s_1 = 1$, $s_2 = 100$, $a_3 = 4$, $s_3 = 1$. Using the default parameters, their shapes are illustrated in Fig. 4 (a), showing that each function enables small velocities to be scaled up by a factor of 10 while maintaining nearly unchanged levels for the other velocities, thereby allowing small velocities to receive greater emphasis in the reconstruction process. It is also noted that the "smoothness" of these four functions decreases in order. The default function used in this study was $f_{U,1}$, which is the smoothest among these functions and is the only monotonic one. Additionally, our SVA strategy can also be considered as a pointwise weighting strategy, where the observed velocities are weighted by $f_U(u) / u$. This weighting is illustrated in Fig. 4 (b), indicating that small and normal velocities are assigned weights of 10 and 1, respectively.

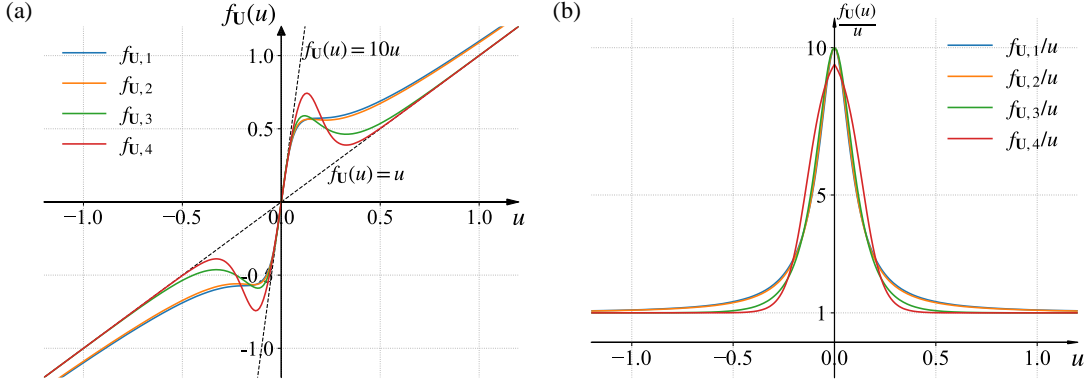

**Fig. 4.** Four transformation functions for the OCs of velocities. (a) Curves of $f_U(u)$. (b) Curves of $f_U(u) / u$.

*2.4 PINN configurations*

The most commonly used NN architecture is the multi-layer perception (MLP); however, we employ a modified MLP (MMLP) [74, 75] with random weight factorization (RWF) [74, 76]. The MLP can be represented by:

$$\begin{aligned}
&\mathbf{a}_0 = \mathbf{x}^{\#} \\
&\mathbf{a}_l = \sigma(\mathbf{z}_l),\ l = 1, 2, ..., L-1 \\
&\mathbf{z}_l = \mathbf{W}_l \mathbf{a}_{l-1} + \mathbf{b}_l,\ l = 1, 2, ..., L \\
&\mathbf{u}_\theta^{\#} = \mathbf{z}_L
\end{aligned} \quad (21)$$

where $\sigma$ is the activation function, $\mathbf{W}$ and $\mathbf{b}$ are weight matrices and biases and are also the trainable parameters ($\boldsymbol{\theta}$). The formulation of MMLP is:

$$\begin{aligned}
&\mathbf{U} = \sigma\left(\mathbf{W}_U \mathbf{x}^{\#} + \mathbf{b}_U\right) \\
&\mathbf{V} = \sigma\left(\mathbf{W}_V \mathbf{x}^{\#} + \mathbf{b}_V\right) \\
&\mathbf{a}_0 = \mathbf{x}^{\#} \\
&\mathbf{a}_l = \sigma(\mathbf{z}_l) \odot \mathbf{U} + (1 - \sigma(\mathbf{z}_l)) \odot \mathbf{V},\ l = 1, 2, ..., L-1 \\
&\mathbf{z}_l = \mathbf{W}_l \mathbf{a}_{l-1} + \mathbf{b}_l,\ l = 1, 2, ..., L \\
&\mathbf{u}_\theta^{\#} = \mathbf{z}_L
\end{aligned} \quad (22)$$

where each hidden layer is reweighted by $\mathbf{U}$ and $\mathbf{V}$, which is similar to the attention mechanism. MMLP has a more complicated structure than MLP, leading to its better performance in various problems [74, 77]. We choose *sin* as the activation function because it is a smooth function and less likely than *tanh* to cause the gradient vanishing. RWF decomposes each weight matrix into a vector ($\mathbf{s}$) and a matrix ($\mathbf{M}$):

$$\mathbf{W} = \mathrm{diag}(\mathbf{s}) \cdot \mathbf{M} \quad (23)$$

Training $\mathbf{s}$ and $\mathbf{M}$ rather than $\mathbf{W}$ can shorten the optimization distance, thus yielding better performance.

As previously stated, PINN samples coordinate points in the spatiotemporal domain to compute PDE residuals, referred to as PDE residual points. Rather than sampling a large quantity of points at once, which is computationally expensive, we sample a small amount of coordinate data and randomly resample every 10 epochs, which can also enhance the model's robustness [78]. The number of resampled PDE residual points is 2000, that is, $|\mathcal{T}_{\text{PDE}}|$ = 2000. For the observed points, the default configuration assumes a uniform distribution over a mesh, whose size is $N_x \times N_y \times N_t = N_{\text{ob}} = |\mathcal{T}_{\text{OC}}|$. The default number of observations is: $N_x$ = 16, $N_x$ = 32, $N_t$ = 51.

In terms of the training configuration, a decaying learning rate strategy is adopted: $\eta_k = 10^{-3} \times r^{\lfloor k/1000 \rfloor}$, where the decay rates ($r$) are 0.98 and 0.97 for variable and constant density cases, respectively. The number of training epochs ($k_{\max}$) is 80000 and 50000 for the respective cases. ADAM [79] is used as the optimizer. The loss weights for PDE ($w_{\text{PDE}}$) are set to 1, while the loss weights for the OCs of $\rho$, $p$, and $c$ are assigned a value of 200. Since the velocities are amplified, their weights ($w_{\text{OC},u}$ and $w_{\text{OC},v}$) are only set to 100. In practice, the loss weights are empirical hyperparameters that need to be tested, and we found the current configuration to be sufficiently effective. The topic of dynamic loss weights is an emerging area of interest [31, 75], which will be considered in our future work.

The reconstruction performance is evaluated on a grid, whose size is 256×512×101 (= $N$). The evaluation metrics we use is the $L_2$ relative error ($L_2$RE):

$$L_2\text{RE}_\varphi = \frac{\|\varphi_{\text{PINN}} - \varphi_{\text{ref}}\|_2}{\|\varphi_{\text{ref}}\|_2} = \frac{\sqrt{\sum_{i=1}^{N}\left(\varphi_{i,\text{PINN}} - \varphi_{i,\text{ref}}\right)^2}}{\sqrt{\sum_{i=1}^{N}\varphi_{i,\text{ref}}^2}} \tag{24}$$

where $\varphi = \rho, u, v, p$, and $c$, and "ref" means the reference value. Our computations were implemented using DeepXDE [80], an open-source PINNs library, and the backend we used was Pytorch.

## 3. Results

In this section, we first present and discuss the overall results of KH-PINN for field reconstruction and parameter inference, providing both quantitative and qualitative results. The performance of KH-PINN and data fitting by NN has also been compared. Subsequently, we validate the effectiveness of the proposed strategies (ME, SVA) and adopted NN configurations. Finally, we discuss the effects of noise, the number of observations, and their distribution.

### 3.1 Overall performance

#### 3.1.1 Field reconstruction

Table 2 presents the evaluation metrics for the KH-PINN results across 20 working conditions, listing the $L_2$RE of PINN-reconstructed fields from a random run. Almost all of the errors are below 2% for variable density cases and 1% for constant density cases, strongly demonstrating the accuracy of KH-PINN in field reconstruction. Typically, the errors for variable density cases are larger than those for constant density cases due to the increased complexity of their fields, which aligns with expectations. For the same reason, the error generally increases with the Reynolds number, particularly in the case of $c$, which is the variable that can best reflect the complexity of the field. It is also noticeable that the errors of $v$ are the largest among the variables for most cases, attributed to the small magnitude of $v$, which results in small denominators of $L_2$RE.

**Table 2**

$L_2$RE of the reconstructed fields by KH-PINN across two density types and ten Reynolds numbers.

| Type | $Re$ | $\rho$ | $u$ | $v$ | $p$ | $c$ |
|---|---|---|---|---|---|---|
| | 1000 | 0.10% | 0.19% | 1.38% | 0.08% | 0.18% |
| | 2000 | 0.19% | 0.27% | 0.67% | 0.10% | 0.38% |
| Variable | 3000 | 0.26% | 0.34% | 0.80% | 0.14% | 0.53% |
| density | 4000 | 0.33% | 0.45% | 1.00% | 0.18% | 0.68% |
| | 5000 | 0.40% | 0.56% | 1.22% | 0.23% | 0.84% |
| | 6000 | 0.49% | 0.66% | 1.45% | 0.28% | 1.04% |

|  | | | | | |
|---|---|---|---|---|---|
| | 7000 | 0.56% | 0.73% | 1.64% | 0.32% | 1.19% |
| | 8000 | 0.67% | 0.84% | 1.91% | 0.36% | 1.44% |
| | 9000 | 0.75% | 0.92% | 2.11% | 0.39% | 1.62% |
| | 10000 | 0.80% | 1.02% | 2.30% | 0.43% | 1.74% |
| | 1000 | | 0.34% | 0.87% | 0.06% | 0.28% |
| | 2000 | | 0.23% | 0.54% | 0.07% | 0.46% |
| | 3000 | | 0.23% | 0.54% | 0.08% | 0.48% |
| | 4000 | | 0.27% | 0.53% | 0.09% | 0.64% |
| Constant density | 5000 | – | 0.29% | 0.59% | 0.10% | 0.78% |
| | 6000 | | 0.32% | 0.61% | 0.11% | 0.86% |
| | 7000 | | 0.30% | 0.59% | 0.11% | 0.92% |
| | 8000 | | 0.34% | 0.65% | 0.12% | 0.99% |
| | 9000 | | 0.34% | 0.63% | 0.13% | 1.13% |
| | 10000 | | 0.36% | 0.68% | 0.14% | 1.23% |

For a visual demonstration, Fig. 5 compares the flow fields reconstructed by KH-PINN with the DNS results, illustrating the vorticity ($\omega$), density ($\rho$), and passive scalar ($c$). The fields of $c$ are not shown for variable density cases, as they are similar to $\rho$; likewise, the fields of $\rho$ are omitted for constant density cases because $\rho$ is not a variable and not predicted by the PINN. The first row shows the results for the lowest Reynolds number in our cases ($Re = 1000$), where the large viscosity leads to a highly dissipative flow. Although not displayed, when $t > 2.5$ s, the vortices are nearly smoothed out and the flow field resembles a double shear layer that gradually widens rather than a rolled-up vortex street. By comparison, the flow fields in the second row for $Re = 5000$ exhibit more convective behavior, showing a noticeable continual roll-up of the vortices, formed by wrapping material from above and below the initial shear layer into thin filaments. During the twisting process of the filaments around the vortices, the stretch-tapering and diffusion effects compete, with the outcome determined by viscosity. The third row also presents the results of $Re = 5000$, but with constant density. Comparing the fields from the second and third rows reveals that vortex centers shift to the right in variable density cases but remain fixed in constant density cases. This behavior is expected, as the momenta above and below each vortex are equal in magnitude and opposite in direction for constant density cases, while the rightward momentum exceeds the leftward momentum in variable density cases. The last row gives the results for the largest Reynolds number in our cases ($Re = 10000$), where the fields are distinctly complex and the filaments are thinner than in the third row, as expected. Each subplot demonstrates excellent agreement between the PINN-reconstructed and DNS-simulated fields for both diffusive and convective flows, as well as for moving and stationary vortices, and for wide and thin filaments. Moreover, the continuity at the centerline indicates the effectiveness of implementing hard periodic BCs. In summary, Table 2 and Fig. 5 can quantitatively and qualitatively validate the accuracy of KH-PINN in reconstructing the flow fields under various conditions, respectively.

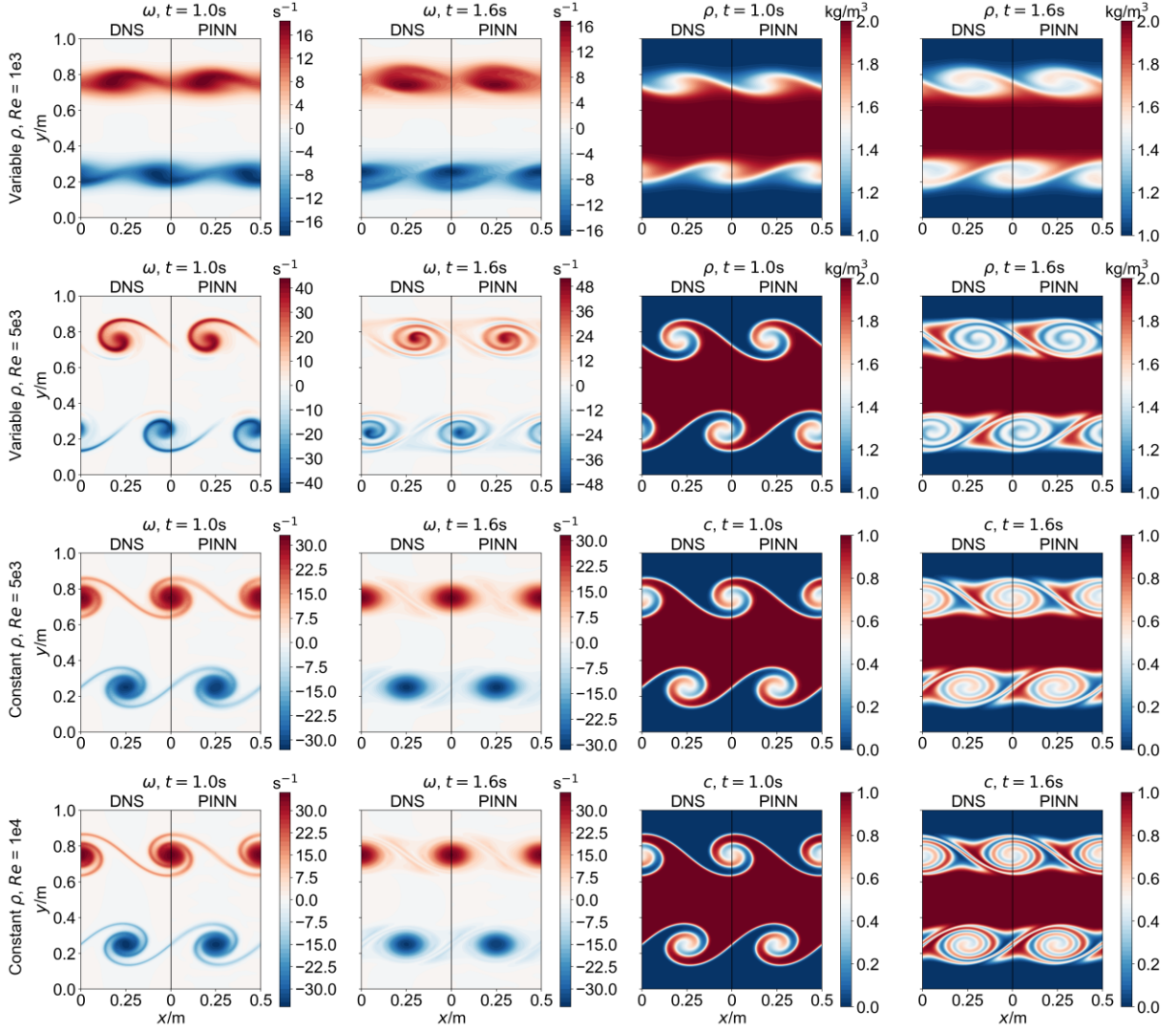

**Fig. 5.** Flow fields reconstructed by KH-PINN at $t = 1.0$ s and $t = 1.6$ s. From the top row to the bottom row, four typical cases are given. For each contour plot, the left and right parts are DNS-simulated and PINN-reconstructed results, respectively.

Fig. 6 gives the curves of mean kinetic energy ($K$) decaying and total scalar entropy ($S$) increasing obtained by DNS and KH-PINN. Their definitions are:

$$K = \int \frac{1}{2}\rho\left(u^2 + v^2\right) dV \Big/ \int 1 dV \tag{25}$$

$$s = -c \ln c, \quad S = \int \rho s \, dV \tag{26}$$

where $s$ is the scalar entropy per unit mass and $V$ represents the volume. Since the flow has no external energy input, the mean kinetic energy will decay with time due to viscous dissipation, and the decaying rate is positively correlated with viscosity. The increase of total scalar entropy is caused by non-reversible dissipation and it can be strictly proved that $dS/dt > 0$ if $D > 0$ [72]. More specifically,

$$\frac{dS}{dt} = \int \rho D \frac{|\nabla c|^2}{c} dV \geq 0 \tag{27}$$

In general, $S$ increases more rapidly with larger $D$, as shown in Fig. 6. However, the increasing rate is also influenced by the distribution of $c$ according to the previous equation, so a larger $D$ may not always yield a larger $S$ at certain time points. Specifically, for $Re \geq 2000$, the entropy curves exhibit the following behavior: (1) $S$ increases slowly when $t < 1$ s, marking the initial development stage of KHI, during which the vortex streets gradually form from the initial double shear layers. (2) $S$ increases rapidly when $1$ s $< t < 2$ s, corresponding to the vortex growth stage characterized by an increasing number of thin filaments twisting around the vortices (see Fig. 5). In this stage, the thin filaments result in many locations with large $\nabla c$, so d$S$/d$t$ is large. (3) $S$ increases slowly when $t > 2$ s, representing the vortex diffusion stage where the widening filaments smooth out the vortices. These three stages collectively create S-shaped curves, which can be considered the normal mode of KHI. In contrast, for $Re = 1000$, $S$ increases almost linearly and no filaments form (as shown in Fig. 5.) due to high viscosity. This flow mode is different from other cases, so at some moment its $S$ is not the largest although its $D$ is the largest. By comparing the results of DNS and KH-PINN, we observe well-matched curves in most cases, further demonstrating the accuracy of KH-PINN. For $Re = 10000$, the reconstructed $K$ deviates slightly from the reference result and $S$ is somewhat overpredicted, but the overall error remains acceptable.

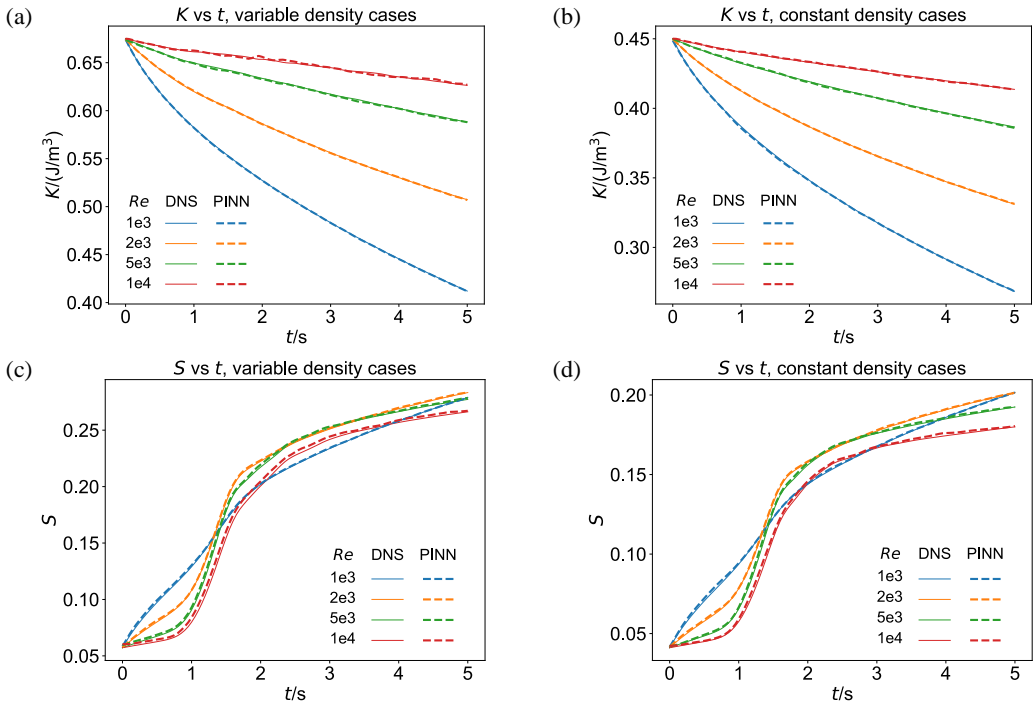

**Fig. 6.** Mean kinetic energy ($K$) and total scalar entropy ($S$) curves obtained by DNS and KH-PINN across two density types and four Reynolds numbers.

To validate the necessity of physical constraint in field reconstruction, we removed the PDE-related loss function terms while keeping all other configurations unchanged, effectively treating the reconstruction as a pure data fitting process. The results are shown in Fig. 7 and Fig. 8. The error curves in Fig. 7 clearly shows that the errors of NN fitting are larger than those of PINN, particularly for $c$ when $1$ s $< t < 2$ s, during which intricate filaments around the vortices are forming. The corresponding contours displayed in Fig. 8 qualitatively shows the inferior performance of NN fitting, which fails to reconstruct even the basic shape of the vortices. The errors are also given in the figure caption, showing substantial values for NN fitting while remaining small for PINNs. In summary, the results robustly demonstrate the necessity and effectiveness of PDE constraints, without which the model exhibits considerably poorer performance in field reconstruction. Moreover, the second limitation of NN fitting without PDE constraints is its inability to infer unknown parameters, a capability that PINN possesses, as detailed in the following section.

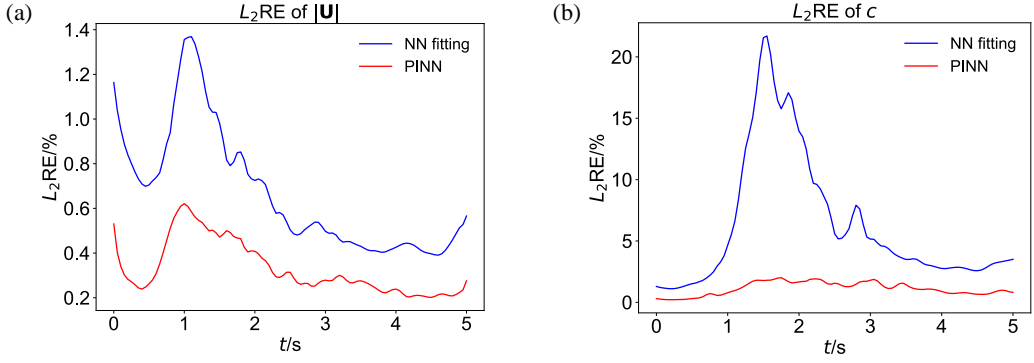

**Fig. 7.** Error curves of the velocity magnitude (|**U**|) and passive scalar (*c*) obtained by PINN and pure NN fitting. Case: Constant density, $Re = 10000$.

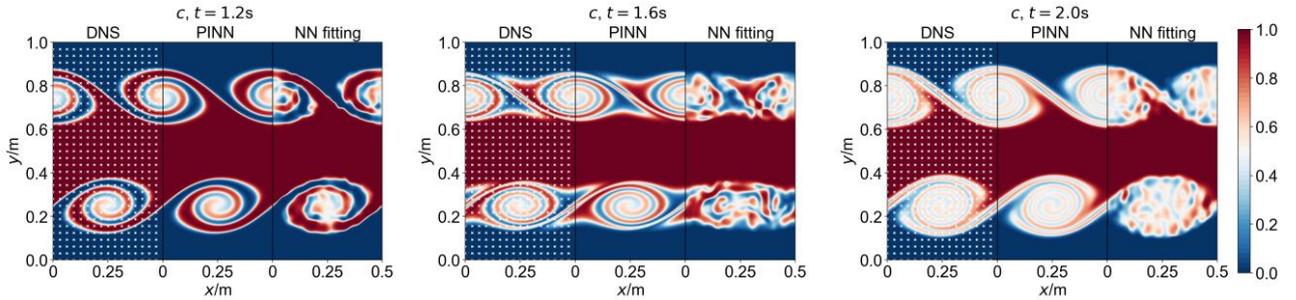

**Fig. 8.** Comparison of the flow fields reconstructed by PINN and pure NN fitting. Case: Constant density, $Re = 10000$. The white points represent the observed points. For $t$ = 1.2 s, 1.6 s, and 2.0 s, the $L_2$REs of PINN results are 1.20%, 1.80%, and 1.66%, respectively, while the $L_2$REs of NN fitting results are 10.58%, 20.22%, and 13.94%, respectively.

*3.1.2 Parameter inference*

Table 3 lists the inferred viscosities and diffusion coefficients by KH-PINN for constant density cases and Fig. 9 presents the learning histories under some Reynolds numbers. The absolute relative errors (AREs) for inferred $\nu$ are less than 10% for most cases, while the AREs for inferred $D$ are remain below 5% for all cases, which demonstrates the accuracy of KH-PINN in parameter inference. When $Re$ is large, the inferred values tend to exceed the true values because the inference is conducted based on discrete observed points, which can naturally result in the overestimation of viscosities and diffusion coefficients when the flow structures are thinner than the observation spacing. Quantitatively, while the ARE for inferred $\nu$ is 17% for $Re = 10000$, the ARE for the initial value is 400%. The accuracy improvement suggests that the results are acceptable. Furthermore, the accuracy could be enhanced by incorporating more observed points. Fig. 9 illustrates that the learning curves exhibit fast convergence and robustness concerning different true values from the same initial values, further highlighting the comprehensive performance of KH-PINN in parameter inference. Although the learning curves for $D$ at $Re = 1000$ deviates slightly from the true value, it can finally converge to the true value with continued training. In summary, the feasibility and accuracy of KH-PINN in parameter inference are validated.

**Table 3**

Results of inferred viscosities and diffusion coefficients by KH-PINN under ten Reynolds numbers for constant density cases. The initial values of $\nu$ and $D$ are all $5\times10^{-4}$ m²/s.

| $Re$ | $\nu$ ($\times10^{-4}$ m²/s) | | $D$ ($\times10^{-4}$ m²/s) | |
|---|---|---|---|---|
| | True | Inferred | True | Inferred |
| 1000 | 10.00 | 9.80 | 10.00 | 10.30 |

| | | | |
|---|---|---|---|
| 2000 | 5.00 | 4.97 | 5.00 | 4.96 |
| 3000 | 3.33 | 3.37 | 3.33 | 3.31 |
| 4000 | 2.50 | 2.59 | 2.50 | 2.49 |
| 5000 | 2.00 | 2.08 | 2.00 | 2.03 |
| 6000 | 1.67 | 1.83 | 1.67 | 1.67 |
| 7000 | 1.43 | 1.55 | 1.43 | 1.45 |
| 8000 | 1.25 | 1.44 | 1.25 | 1.29 |
| 9000 | 1.11 | 1.24 | 1.11 | 1.16 |
| 10000 | 1.00 | 1.17 | 1.00 | 1.04 |

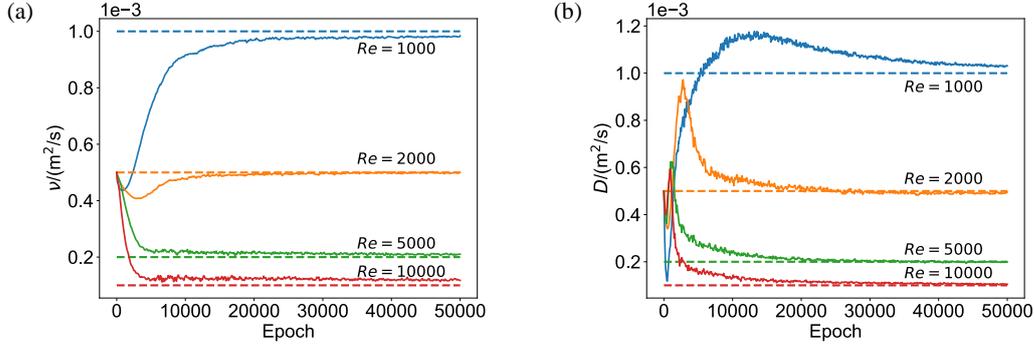

**Fig. 9.** Parameter learning histories under four Reynolds numbers for constant density cases. (a) Viscosity. (b) Diffusion coefficient.

### 3.2 Strategy and hyperparameter validation

#### 3.2.1 Multiscale embedding (ME)

Table 4 lists the errors of reconstructed fields using different input embedding strategies and Fig. 10 further compares the results of KH-PINN with and without ME. In Table 4, "Fourier" indicates that Eq. (16) is used as the input embedding, whereas "linear" indicates that $x^\# = [x^*, 2x^*, \ldots, nx^*]$; similar conventions apply for $y^\#$ and $t^\#$. The bold texts represent the default configuration we adopted, which yields the best performance. More discussions are as follows. (1) The first two rows are results without any MEs, which are inferior to most cases in the table, underscoring the necessity of ME. Moreover, the first row is the results with fully linear input embeddings where the hard constraint of periodic BCs is absent, resulting in the expected worst performance. (2) The third and fourth rows are results with only spatial MEs, which perform worse than those with both spatial and temporal MEs, indicating the importance of temporal ME. The results are worse than the second row for $u$, $v$, and $p$, but better for $c$ because $c$ exhibits more multiscale features than other variables. (3) The fifth and sixth rows show results with nonlinear spatial MEs and linear temporal MEs, which are superior to the third and fourth rows but inferior to the eighth and ninth rows. This suggests that temporal MEs are effective and that nonlinear MEs are more effective than linear ones. (4) The last four rows reveal that there is an optimal embedding order ($n$), as a small $n$ provides insufficient multiscale features while a large $n$ poses challenges for the network in learning effective scales. Currently, $n$ remains a hyperparameter in ongoing studies [43, 52, 54], but it can be set more reasonably if the multiscale information of the problem is approximately known in advance. In summary, Table 4 can quantitatively validate the effectiveness of the ME strategy and demonstrates the advantage of our ME configurations over others.

**Table 4**

$L_2$RE of the reconstructed fields by KH-PINN using different input embedding strategies. Case: Constant density, $Re = 10000$. $n$ is the embedding order and $n = 1$ means no multiscale embedding. The bold texts represent the default configuration we adopted.

| $x, y$ | $t$ | $u$ | $v$ | $p$ | $c$ |
|---|---|---|---|---|---|
| Linear, $n = 1$ | Linear, $n = 1$ | 0.92% | 1.86% | 0.36% | 2.75% |
| Fourier, $n = 1$ | Linear, $n = 1$ | 0.48% | 0.91% | 0.18% | 1.63% |
| Fourier, $n = 4$ | Linear, $n = 1$ | 0.52% | 0.94% | 0.21% | 1.55% |
| Fourier, $n = 8$ | Linear, $n = 1$ | 0.60% | 0.97% | 0.24% | 1.58% |
| Fourier, $n = 4$ | Linear, $n = 4$ | 0.44% | 0.76% | 0.17% | 1.38% |
| Fourier, $n = 8$ | Linear, $n = 8$ | 0.52% | 0.86% | 0.19% | 1.60% |
| Fourier, $n = 2$ | Fourier, $n = 2$ | 0.39% | 0.70% | 0.15% | 1.28% |
| **Fourier, $n = 4$** | **Fourier, $n = 4$** | **0.36%** | **0.68%** | **0.14%** | **1.23%** |
| Fourier, $n = 8$ | Fourier, $n = 8$ | 0.45% | 0.75% | 0.14% | 1.38% |
| Fourier, $n = 16$ | Fourier, $n = 16$ | 0.50% | 0.83% | 0.16% | 1.40% |

In Fig. 10, "no ME" and "with ME" correspond to the second and eighth rows in Table 4, respectively. Fig. 10 (a) clearly shows that the overall errors with ME are lower than those without ME. Fig. 10 (b) compares the fields of $c$ at $t = 2.0$ s. Although the differences between the results with and without ME are not visually significant, their $L_2$REs are 1.66% and 2.43%, respectively. To show the differences more clearly, Fig. 10 (c) and (d) give the corresponding curve plots, which show that when high frequency (small scales) occurs, PINN cannot accurately reconstruct the fields without ME, while the results are satisfactory when ME is employed. Fig. 10 (e) and (f) display TKE spectra at two moments, revealing that the small-scale energies without ME are lower than those with ME, which is as expected. In summary, Fig. 10 further qualitatively demonstrates the effectiveness of the proposed ME strategy.

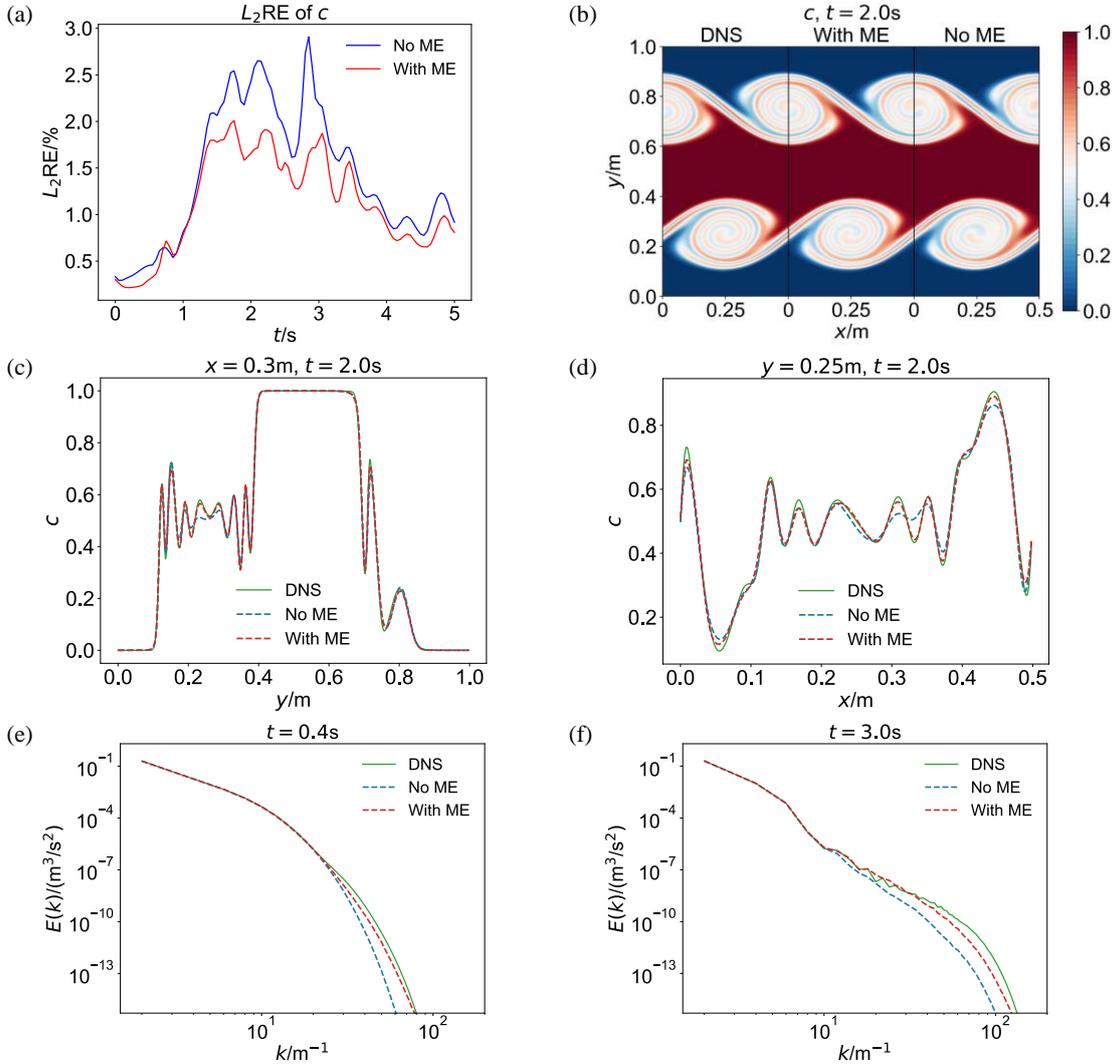

**Fig. 10.** Comparison between the results of KH-PINN with and without multiscale embedding (ME). Case: Constant density, $Re = 10000$. (a) Error curves of reconstructed $c$. (b) Reconstructed $c$ at $t = 2.0$ s. The $L_2$REs of the 2D fields are 1.66% and 2.43% with and without ME, respectively. (c) Reconstructed $c$ at $t = 2.0$ s and $x = 0.3$ m. (d) Reconstructed $c$ at $t = 2.0$ s and $y = 0.25$ m. (e) and (f) are TKE spectra at $t = 0.4$ s and $t = 3.0$ s, respectively.

### 3.2.2 Small-velocity amplification (SVA)

Table 5 lists the errors of reconstructed fields using various velocity transformation functions ($f_U$). The performances of $f_{U,1}$ and $f_{U,2}$ are comparable and satisfactory. In contrast, $f_{U,3}$ and $f_{U,4}$ yield significantly poorer results due to their non-monotonic nature, leading to a situation where the same $f_U(u)$ can correspond to multiple values of $u$. Consequently, even when $f_U(u)$ is accurately approximated, the reconstructed velocity may still be erroneous. Therefore, $f_{U,1}$ is the most highly recommended option for practical applications. A comparison of the results in the first two rows indicates that while SVA can increase the mean errors, the overall results remain satisfactory. More importantly, SVA enhances the reconstruction accuracy for small velocities, which is particularly significant in KHI problems. To further elucidate this point, Fig. 11 compares the reconstructed small $v$ of KH-PINN with and without SVA. Two typical types of small velocities are shown: The small velocity perturbations that induce the KHI ($t = 0.0$ s) and the small velocities formed during the late dissipation stage ($t = 4.9$ s). The results in (a) to (f) demonstrate that PINN can accurately reconstruct these two types of small velocities when SVA is applied, but fails to do so in its absence, thereby robustly validating the effectiveness of the proposed SVA strategy.

**Table 5**

$L_2$RE of the reconstructed fields by KH-PINN using various velocity transformation functions. Case: Constant density, $Re = 10000$. The bold texts represent the default configuration we adopted.

| $f_U$ | $u$ | $v$ | $p$ | $c$ |
| --- | --- | --- | --- | --- |
| No | 0.279% | 0.453% | 0.119% | 1.144% |
| **$f_{U,1}$** | **0.357%** | **0.683%** | **0.144%** | **1.227%** |
| $f_{U,2}$ | 0.357% | 0.683% | 0.141% | 1.272% |
| $f_{U,3}$ | 87.17% | 62.94% | 22.55% | 8.70% |
| $f_{U,4}$ | 80.45% | 66.51% | 20.81% | 9.16% |

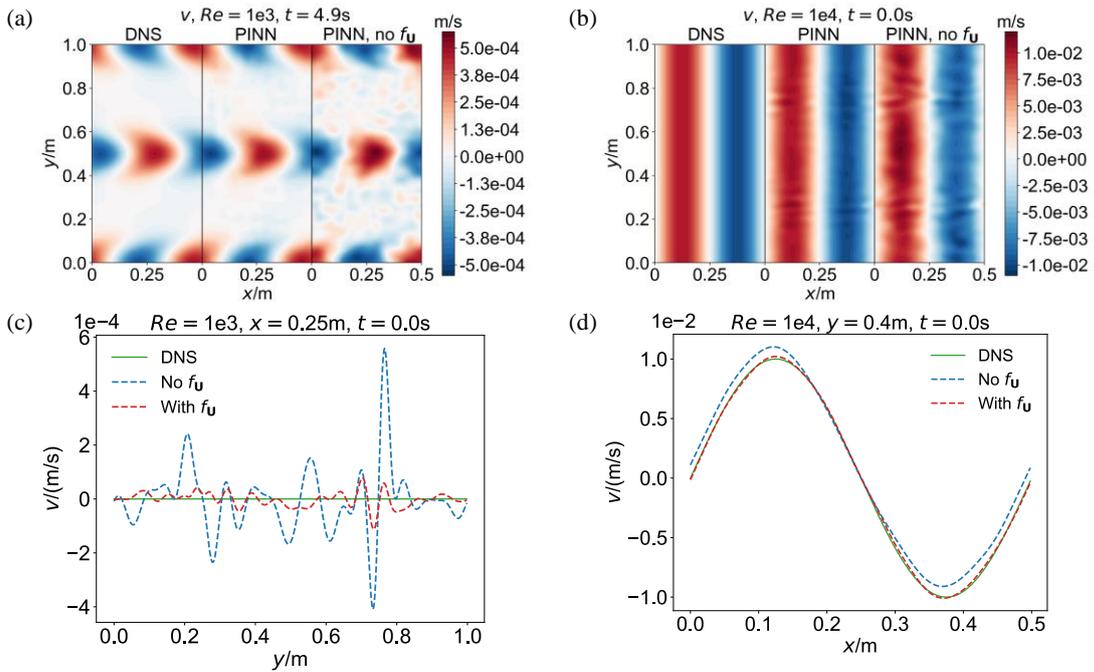

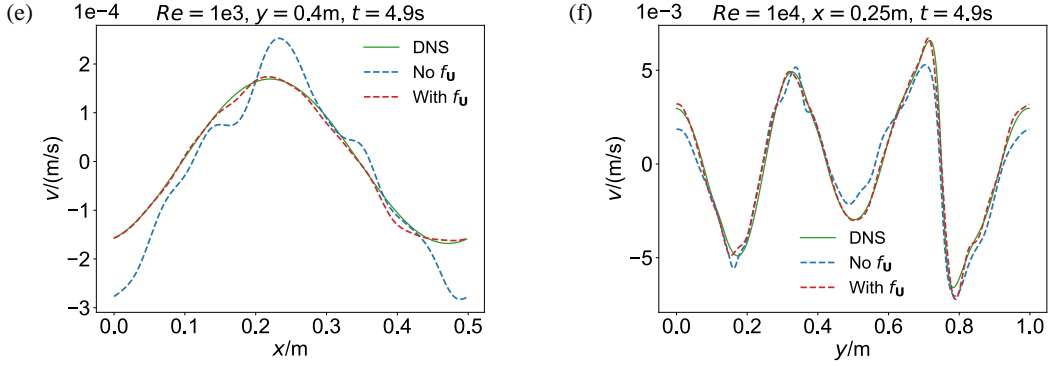

**Fig. 11.** Comparison between the reconstructed small $v$ of KH-PINN with and without small-velocity amplification strategy. Case: Constant density, $Re$ = 10000. Typical $Re$ and moments are chosen. (a) and (b) are 2D fields, with the $L_2$REs being 6.62% and 3.36% for the results with $f_U$, respectively, and being 30.36% and 12.05% for the results without $f_U$, respectively. (c) to (f) are 1D distributions at specified $x$ and $y$.

*3.2.3 Network configurations*

Table 6 lists the errors of reconstructed fields with various network configurations. A comparison of the first four rows reveals that both MMLP and RWF are effective in enhancing the reconstruction accuracy, with MMLP demonstrating superior performance compared to RWF. Fig. 12 compares the results using MLP and MMLP-RWF, where (a) and (b) highlight the limitations of MLP in learning high-frequency (small-scale) features. Fig. 12 (c) also indicates that the small-scale energies obtained using MLP are lower than those derived from MMLP-RWF and DNS. Fig. 12 (d) illustrates that MMLP-RWF outperforms MLP in reconstructing velocity fields. In short, the effectiveness of both MMLP and RWF has been validated quantitatively and qualitatively. Moreover, the fourth row of Table 6 suggests that *sin* is more suitable as the activation function for our problem than *tanh*. This preference may stem from the oscillatory nature of *sin*, which aligns better with the high-frequency characteristics of our problem, consistent with the conclusions of other studies [81, 82].

**Table 6**
$L_2$RE of the reconstructed fields by KH-PINN with various network configurations. Case: Constant density, $Re$ = 10000. The bold texts represent the default configuration we adopted.

| NN | RWF | $\sigma$ | $u$ | $v$ | $p$ | $c$ |
| --- | --- | --- | --- | --- | --- | --- |
| **MMLP** | **Yes** | **sin** | **0.36%** | **0.68%** | **0.14%** | **1.23%** |
| MMLP | No | sin | 0.42% | 0.81% | 0.16% | 1.62% |
| MLP | Yes | sin | 0.78% | 1.88% | 0.27% | 2.40% |
| MLP | No | sin | 0.93% | 2.21% | 0.32% | 3.03% |
| MMLP | Yes | tanh | 0.39% | 0.73% | 0.15% | 1.36% |

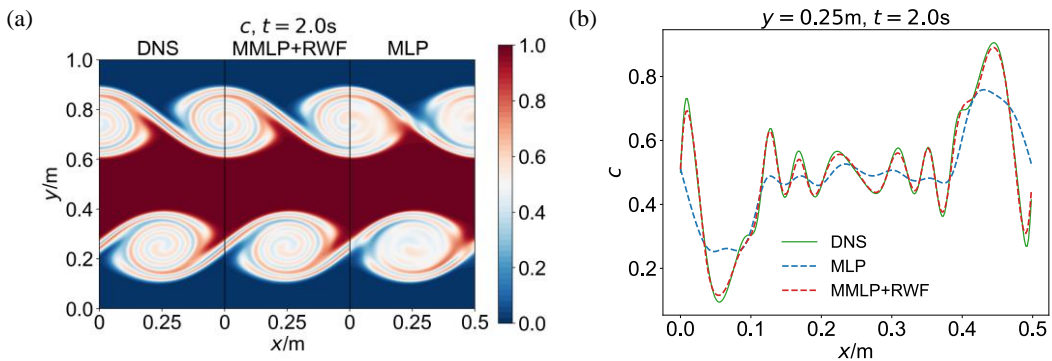

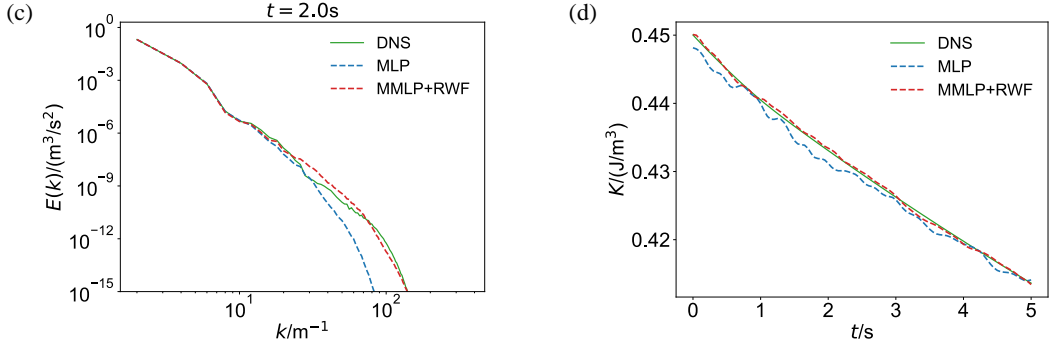

**Fig. 12.** Comparison between the results of KH-PINN using MLP and MMLP-RWF. Case: Constant density, $Re = 10000$. (a) Reconstructed $c$ at $t = 2.0$ s. The $L_2$REs of the 2D fields are 1.66% and 6.68% using MMLP-RWF and MLP, respectively. (b) Reconstructed $c$ at $t = 2.0$ s and $y = 0.25$ m. (c) TKE spectra at $t = 2.0$ s. (d) Mean kinetic energy decaying curves.

*3.2.4 Observation configurations*

Table 7 lists the errors of reconstructed fields at different observation noise levels and Fig. 13 gives the results of some noise levels. Consistent with expectations, the accuracy of reconstruction deteriorates as the noise level increases. However, both Table 7 and Fig. 13 indicate that satisfactory results are achieved when the noise level is below 10% and that the basic shapes of the flow field can be effectively reconstructed even at noise levels as high as 20%. Thus, the anti-noise capability of KH-PINN is validated.

**Table 7**

$L_2$RE of the reconstructed fields by KH-PINN at different observation noise levels. Case: Constant density, $Re = 5000$. The bold texts represent the default configuration we adopted.

| Noise | $u$ | $v$ | $p$ | $c$ |
| --- | --- | --- | --- | --- |
| **0%** | **0.29%** | **0.59%** | **0.10%** | **0.78%** |
| 1% | 0.32% | 0.62% | 0.13% | 0.77% |
| 2% | 0.39% | 0.80% | 0.20% | 0.94% |
| 5% | 0.70% | 1.55% | 0.41% | 1.62% |
| 10% | 1.33% | 2.93% | 0.69% | 3.02% |
| 15% | 1.84% | 4.45% | 0.93% | 4.44% |
| 20% | 2.44% | 5.51% | 1.15% | 5.93% |

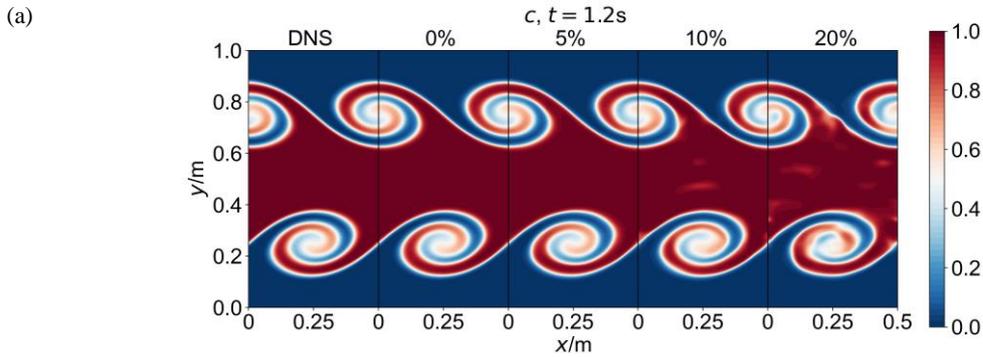

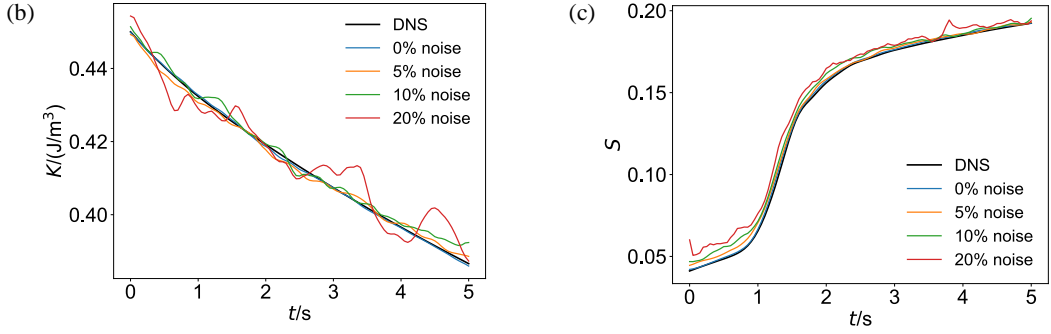

**Fig. 13.** Results of KH-PINN at different noise levels. Case: Constant density, $Re = 5000$. (a) Reconstructed $c$ at $t = 1.2$ s. The $L_2$REs of the 2D fields are 0.99%, 2.06%, 3.50%, and 6.58% for 0%, 5%, 10%, and 20% noise levels, respectively. (b) Mean kinetic energy decaying curves. (c) Total scalar entropy increasing curves.

Table 8 lists the errors of reconstructed fields with different numbers ($N_{ob}$) and distributions of observed points. Generally, the errors decrease as $N_{ob}$ increases. Other discussions are as follows. (1) The first six rows indicate that increasing the number of observations to more than double the default configuration yields only marginal improvements. (2) The temporal resolution of observations should not be excessively low; otherwise, the result will be bad. Specifically, results are satisfactory when $N_t \geq 26$. This corresponds to a temporal resolution of less than 0.2 s, which is relatively low in practical applications. (3) When $N_t \geq 26$ and with equal $N_t$ and $N_{ob}$, the errors remain similar across various $N_x$ and $N_y$ values, suggesting that the spatial distribution of the observations has minimal impact on model performance in this context. (4) When $N_t \geq 26$, despite reducing the number of observations to eight times the default configuration, results remain satisfactory, strongly demonstrating the few-shot learning capability of our KH-PINN.

**Table 8**

$L_2$RE of the reconstructed fields by KH-PINN with different numbers and distributions of observed points. Case: Constant density, $Re = 5000$. The bold texts represent the default configuration we adopted.

| Factor | $N_x, N_y, N_t$ | $u$ | $v$ | $p$ | $c$ |
|---|---|---|---|---|---|
| 4 | 32, 64, 51 | 0.24% | 0.49% | 0.09% | 0.62% |
|  | 32 32, 101 | 0.27% | 0.47% | 0.11% | 0.65% |
|  | 16, 64, 101 | 0.24% | 0.49% | 0.10% | 0.66% |
| 2 | 32, 32, 51 | 0.26% | 0.50% | 0.10% | 0.66% |
|  | 16, 64, 51 | 0.26% | 0.54% | 0.10% | 0.66% |
|  | 16, 32, 101 | 0.30% | 0.55% | 0.11% | 0.76% |
| 1 | **16, 32, 51** | **0.29%** | **0.59%** | **0.10%** | **0.78%** |
|  | 32, 16, 51 | 1.18% | 0.54% | 0.10% | 0.79% |
|  | 32, 32, 26 | 0.30% | 0.70% | 0.12% | 0.71% |
|  | 32, 64, 13 | 1.22% | 6.02% | 1.00% | 6.31% |
| 0.5 | 8, 32, 51 | 0.34% | 0.77% | 0.12% | 1.05% |
|  | 16, 16, 51 | 1.24% | 0.66% | 0.11% | 1.02% |
|  | 16, 32, 26 | 0.34% | 0.80% | 0.12% | 0.83% |
| 0.25 | 8, 16, 51 | 1.38% | 0.86% | 0.12% | 1.72% |
|  | 8, 32, 26 | 0.36% | 1.04% | 0.14% | 1.25% |
|  | 16, 16, 26 | 1.04% | 0.87% | 0.12% | 1.17% |
|  | 16, 32, 13 | 1.22% | 6.02% | 1.02% | 6.23% |
| 0.125 | 8, 8, 51 | 1.75% | 1.51% | 0.16% | 2.98% |
|  | 4, 16, 51 | 1.96% | 1.44% | 0.20% | 3.09% |
|  | 8, 16, 26 | 1.18% | 1.15% | 0.14% | 1.76% |
|  | 16, 16, 13 | 1.86% | 6.02% | 1.06% | 5.93% |
|  | 16, 32, 7 | 1.38% | 6.66% | 3.91% | 5.96% |

## 4. Conclusions

In this study, the KH-PINN framework is established to solve inverse problems of KHI based on PINNs. The 2D unsteady incompressible KHI is studied across a broad range of Reynolds numbers, considering both constant and variable densities. Field reconstruction and parameter inference are performed based on sparse observations. To tackle the challenges associated with spatiotemporal and magnitude multiscale issues, the strategies of multiscale embedding (ME) and small-velocity amplification (SVA) are proposed. The main conclusions are:

(1) KH-PINN can accurately reconstruct the flow fields and infer the transport parameters across various working conditions. It effectively reconstructs dissipative and convective flows, stationary and moving vortices, as well as normal-scale main flows and small-scale filaments. The obtained mean kinetic energy decaying and total scalar entropy increasing curves closely align with the reference ones. Furthermore, KH-PINN significantly outperforms pure data fitting using NNs, underscoring the efficacy of incorporating physical PDE constraints.

(2) It is validated that the adopted ME strategy enhances the reconstruction accuracy for small-scale structures, and the proposed SVA strategy improves the reconstruction accuracy for small-magnitude velocities, which is important for KHI problems. The ME strategy can be extended to other problems with spatiotemporal multiscale, and the SVA strategy can be extended to scenarios where small-magnitude variables are important yet challenging to learn. An example is combustion problems, where numerous minor species, such as NOx and radicals, serve as important variables.

(3) The effectiveness of the adopted NN architecture has been validated. Moreover, by changing the observation configurations, we have further confirmed the anti-noise and few-shot learning capabilities of KH-PINN, thereby enhancing its applicability to more practical scenarios.

To further explore the capability and broaden the applicability of KH-PINN, potential future work includes: (1) Testing cases with higher Reynolds numbers where the spatiotemporal multiscale is more significant. (2) Developing KH-PINN for compressible KHI flows, which exhibit greater nonlinear characteristics than incompressible flows. (3) Applying KH-PINN to practical scenarios where experimental data are available. (4) Extending KH-PINN to create surrogate models by parameterizing the PDEs and NNs, enabling simultaneous learning across multiple working conditions and facilitating rapid predictions of flow fields under varying conditions.

**Funding**

This work is supported by the National Key R&D Program of China (NO. 2021YFA0716202) and the National Natural Science Foundation of China (Grant No. 62273149).

**Declaration of competing interest**

The authors declare that they have no known competing financial interests or personal relationships that could have appeared to influence the work reported in this paper.